\documentstyle[aps,amsmath,amssymb,twocolumn,graphics]{revtex}
\tighten
\begin{document}
\draft
\title{Anisotropic Scattering Rates 
       in the $\mbox{\boldmath$t-t'-U$}$ Hubbard Model} 
\author{Joachim Altmann$^1$, Wolfram Brenig$^1$, and Arno P. Kampf$^2$} 
\address{$^1$ Institut f\"ur Theoretische Physik, 
Universit\"at zu K\"oln,\\ Z\"ulpicher Str. 77, 50937 K\"oln, Germany}
\address{$^2$ Theoretische Physik III, Elektronische Korrelationen und 
Magnetismus,\\ Universit\"at Augsburg, D--86135 Augsburg, Germany}
\address{~ \parbox{14cm}{\rm  \medskip We have investigated the evolution of 
the electronic properties of the $t$-$t$'-$U$ Hubbard model with hole doping 
and temperature. Due to the shape of the Fermi surface, scattering from short 
wavelength spin fluctuations leads to strongly anisotropic quasi--particle 
scattering rates at low temperatures near half--filling. As a consequence, 
significant variations with momenta near the Fermi surface emerge for the 
spectral functions and the corresponding ARPES signals. This behavior is quite 
in contrast to the intermediate doping regime and we discuss the possible 
relevance of our results for the interpretation of photoemission spectra in 
cuprate superconductors at different hole doping levels.}} 
\maketitle \narrowtext 
A key issue in the efforts to understand the microscopic physics of
high--$T_c$ superconductors is the evolution of the electronic
properties with doping. In recent years theoretical work has
continuously benefited from angular resolved photoemission
spectroscopy (ARPES) data for the electronic spectrum and the Fermi
surface in the normal state as well as for the anisotropic energy gap in the
superconducting state \cite{ShenDessau}. In particular, remarkable
recent ARPES results for the underdoped cuprates have shown an
anisotropic normal--state pseudogap which forms below $\sim 150$K for
the weakly underdoped materials increasing up to $\sim 300$K for the
heavily underdoped compounds with a $T_C$ close to zero
\cite{Marshall,Loeser,Ding1,Ding2}.  Contrary to optimally doped and
overdoped samples the quasi--particle peak in the underdoped spectra
is found to be very weak near the $(\pi,0)$ point of the Brillouin zone (BZ)
and no Fermi surface crossing is observed on the BZ boundary along the 
$(\pi,0)$ to $(\pi,\pi)$ direction. Furthermore the spin susceptibility
\cite{Oda}, $c$--axis optical \cite{Homes93a} as well as in--plane infrared
conductivity \cite{Basov}, NMR relaxation rates \cite{Takigawa}, and
inelastic neutron scattering data \cite{Rossat91a} indicate a
pseudogap in the low--energy excitation spectrum of underdoped
compounds. Different scenarios like pair formation well above $T_c$
\cite{Doniach90aUemura91aRanderia92a,EmeryRanningerLevin},
spin--charge separation \cite{Fukuyama92a,Lee} or
precursor  effects near the antiferromagnetic (AF) instability
\cite{ShenSchrieffer,PinesChubukov} have been proposed as possible
origins of these pseudogap phenomena. 

In this paper, we explore the combined effects of strong spin
fluctuation scattering and Fermi surface topology using the 2D Hubbard model 
on a square lattice with a $t-t'$ dispersion of the one--particle
kinetic energy
\begin{equation}\label{1}
\makebox[-.1cm]{}
\epsilon_{\bf k}=-2t(\cos{k_x}+\cos{k_y})-4t'\cos{k_x}\cos{k_y} 
\end{equation}
with nearest--neighbor ($t$) and next--nearest neighbor ($t'$) hopping 
amplitudes. Near half--filling we
demonstrate that strong quasi--particle  scattering rates develop with
decreasing temperature near the so called ``hot  spots'' on the Fermi
surface (FS), i.e. FS points which are connected by the AF wave vectors
${\bf Q}=(\pm\pi,\pm\pi)$. As a consequence of the emerging highly 
anisotropic scattering rates the quasi--particle peaks in the spectral
functions  near the $(\pi,0)$ points of the BZ are suppressed in
comparison to the momenta near ${\bf k}_F$ along the BZ diagonal leading
to highly anisotropic ARPES signals.

Our method of choice to evaluate the renormalized one--particle excitations
is the self--consistent and conserving \cite{Baym62} fluctuation--exchange
(FLEX) approximation \cite{Bickers89a}. In this approach the self--energy
is given in terms of the spin-- and density--fluctuation $T$--matrices 
$T_{sf}({\bf r},\tau)$ and $T_{\rho\rho}({\bf r},\tau)$ by
\begin{equation}\label{2}
\makebox[-.1cm]{}
\Sigma ({\bf r},\tau) =
U^2 G({\bf r},\tau) [\chi_0 ({\bf r},\tau) + T_{\rho\rho}({\bf r},\tau) +
T_{sf}({\bf r},\tau) ]
\end{equation}
where $\chi_0({\bf r},\tau)=-G({\bf r},\tau)G(-{\bf r},-\tau)$ is the 
particle--hole bubble and ${\bf r}$ and $\tau$ denote real space 
coordinates and imaginary time, respectively. $U$ is the on--site Coulomb 
repulsion. The Fourier transformed 
$T_{sf}({\bf q},{\rm i}\omega_m)$ and $T_{\rho\rho}({\bf q},{\rm i}\omega_m)$ 
are defined by
\begin{eqnarray}\label{3}
 T_{\rho\rho} ({\bf q},{\rm i}\omega_m) = -\frac{1}{2} \frac{U
\chi_0^2 ({\bf q},{\rm i}\omega_m)} {1 + U \chi_0 ({\bf q},{\rm i}\omega_m)}
\\ \label{4}
 T_{sf}({\bf q},{\rm i}\omega_m) = \frac{3}{2} \frac{U \chi_0^2 ({\bf q},
{\rm i}\omega_m)} {1 - U \chi_0 ({\bf q},{\rm i}\omega_m)}
\end{eqnarray}
where $\omega_m=2m\pi T$ are the bosonic Matsubara frequencies at temperature 
$T$. In combination with Dyson's equation $G^{-1}=G_0^{-1}-\Sigma$, Eqs. 
(\ref{1}), (\ref{2}), and (\ref{3}) form a self--consistent set of equations 
which we solve numerically by iteration. One of the keys to the numerical 
solution is the {\em locality} of (\ref{2}) in space and time as well as the 
{\em locality} of (\ref{3}) and (\ref{4}) in momentum and frequency space. 
This allows for a completely algebraic treatment of the FLEX iterations by 
repeated application of intermediate Fast--Fourier--transforms \cite{Hess92a}. 
The proper {\em stability} of the self--consistent cycle is achieved by 
solving the FLEX equations on a contour in the complex frequency plane which 
is shifted off the real axis by a finite amount ${\rm i}\gamma$ with 
$0<\gamma< \pi T/2$ \cite{Schmalian96a}. Analytic continuation to the real 
frequency axis does not encounter the usual problems of purely imaginary 
frequency methods \cite{Hess92a}.

We have solved the FLEX equations on lattices with up to 128$\times$128 sites 
using an equally spaced frequency mesh of 4096 points within an energy
window of $[-30t,30t]$. The lower bound on the temperature accessible in
our present calculations is $T\sim 0.02t$. This bound is set by the smallest 
width in frequency space of the AF paramagnon--peak in Im 
$T_{sf}({\bf q},\omega+ {\rm i}\gamma)$ which can be resolved for the
chosen frequency mesh as well as the smallest momentum space width which
can be treated without introducing finite size effects. Throughout the
paper we will adopt an interaction strength $U=4t$ and $t'=-0.3t$ -- a
parameter set for which the Hubbard model exhibits long range AF order
in the ground state at half--filling \cite{Duffy95}.

\setlength{\unitlength}{1cm}
\begin{figure}
\begin{picture}(8.7,7)(1.3,6.8)
  \put(0,0){\scalebox{0.5}{\includegraphics{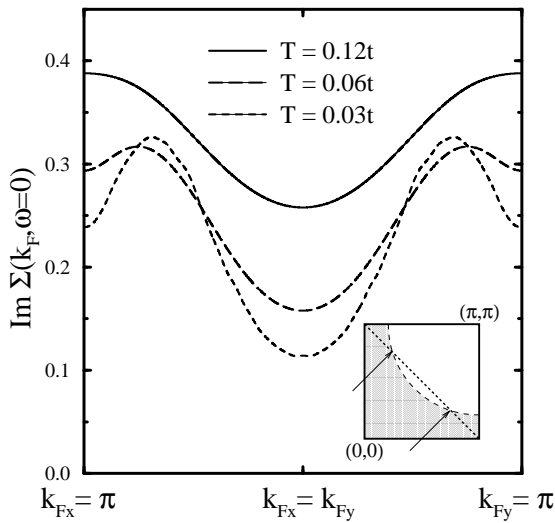}}}
\end{picture}
\caption{Im $\Sigma({\bf k}_F,\omega=0)$ along the Fermi line for various
  temperatures $T$ with $U\! =\! 4t$, $t'\! =\! -0.3t$ and hole
  doping $\delta \! = \! 2.5\%$. The inset shows schematically the
  Fermi line (dashed line) and the ``hot spots'' (arrows) in one
  quarter of the BZ.}
\label{fig1}
\end{figure}

We start with the discussion of our results by considering the FS
anisotropy of the one--particle scattering rate for small hole
concentrations. Its momentum dependence is governed by two effects: First,
the available recoil phase--space which is linked to the momentum--space
width of the AF paramagnon peak in Im $T_{sf}({\bf q},\omega)$ and second
the density of intermediate states. While the former quantity exhibits a
variation with temperature and doping which discriminates only weakly
between the $t$-- and the $t$--$t'$ model in the low doping limit, the
latter quantity depends crucially on the $t-t'$ band structure. This is a
major source of difference between the one--particle renormalizations in
the $t$-- and $t$--$t'$ Hubbard models. In particular, we observe a
non--trivial temperature and momentum dependence of the one--particle
self--energy. This is shown in Fig. \ref{1} which depicts the imaginary part
of $\Sigma({\bf k},\omega=0)$ along the FS line, i.e. at ${\bf k}={\bf
k}_F$. For $T\gtrsim 0.08t$ the AF peak in Im $T_{sf}({\bf q},\omega)$ is a
relatively broad structure, i.e. its HWFM in momentum and frequency space
is $\sim\pi/6$ and $\sim 0.2t$, respectively. Therefore, in essence,
$\Sigma({\bf k}_F,\omega=0)$ is modulated only by the density of states
along the FS which is largest at the borders of the BZ and smallest at the
point ${\bf k}_{F\gamma}$ on the BZ diagonal, i.e.  where
$k_{Fx}=k_{Fy}$. For decreasing temperatures the peak in Im $T_{sf}({\bf
q},\omega)$ at the AF wave vector ${\bf q}={\bf Q}$ sharpens until we loose
its accurate resolution at about $T\approx 0.02t$. This redistribution of
weight shifts the maximum in Im $\Sigma({\bf k}_F,\omega=0)$ into the
so--called ``hot spots'' on the FS which can be connected by the AF wave
vector. In addition to this shift the low--temperature anisotropy of the
self--energy is enhanced by roughly a factor of two.

\begin{figure}
\begin{picture}(8.7,9)(2.4,17.8)
\put(0,0){\includegraphics{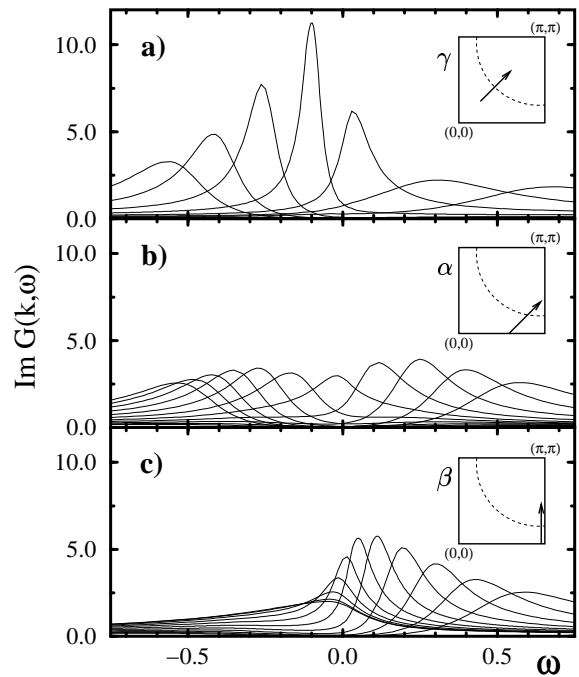}}
\end{picture}
\caption{Im $G({\bf k},\omega)$ along three different paths in the BZ -- 
  schematically shown in the insets -- for $T \! =\! 0.03 t$, $U\! =\! 4t$, 
  $t'\! =\! -0.3t$ and $\delta\! =\! 2.5\%$ on a 64$\times$64 lattice. The 
  path $\gamma$ in a) is chosen along the BZ diagonal from ${\bf k}_\gamma=
  (11,11)\frac{\pi}{32}$ to $(17,17)\frac{\pi}{32}$. In b) the path $\alpha$ 
  is parallel to the path in a) but runs from ${\bf k}_\alpha=(21,0)\frac{\pi}
  {32}$ to $(32,11)\frac{\pi}{32}$ crossing the (hot spot) maximum of Im 
  $\Sigma({\bf k}_F,\omega\! =\! 0)$ at ${\bf k}_{F\alpha}\approx(26,5)
  \frac{\pi}{32}$. The path $\beta$ in c) is along the BZ boundary from ${\bf 
  k}_\beta=(32,0)\frac{\pi}{32}$ to ${\bf k}_\beta=(32,11)\frac{\pi}{32}$.}
\label{fig2}
\end{figure}

In Fig. \ref{fig2} we show the consequence of the anisotropic scattering rates 
for the single--particle spectral function at $T=0.03t$. Choosing a path in 
momentum space as shown in Fig. \ref{fig2}a) which crosses the FS at 
${\bf k}_{F\gamma}$, i.e. the wave vector of the minimal scattering rate, 
a well defined, sharp quasi--particle peak is observed. In contrast, the 
quasi--particle feature is strongly suppressed in Fig. \ref{fig2}b) where the 
FS is traversed by passing through a hot spot. In comparison to Fig. 
\ref{fig2}a) the amplitude of the quasi--particle peak near ${\bf k}_F$ is 
reduced by almost a factor of three and, moreover, it is {\em minimal on the 
FS}. Finally, choosing a path which cuts the FS on the BZ boundary as in Fig. 
\ref{fig2}c), the quasi--particle structure is asymmetrically distributed as a 
function of momentum being more pronounced in the inverse photoemission sector.
Only weak dispersion of the quasi--particle peak below the Fermi energy along 
this cut signals a ``flat--band'' region close to the momentum $(\pi,0)$. We 
note that ARPES spectra can be obtained from Fig. \ref{fig2}a)--c) by 
multiplying Im $G({\bf k},\omega)$ with the Fermi function. Only near the 
momentum ${\bf k}_{F\gamma}$ these spectra are found to display well defined 
quasi--particles which disperse through the Fermi energy, while the 
quasi--particle weight in the vicinity of the hot spots and the BZ boundary
is substantially reduced.

\begin{figure}
\begin{picture}(8.7,11.5)(1.2,4.7)
\put(0,0){\scalebox{0.58}{\includegraphics{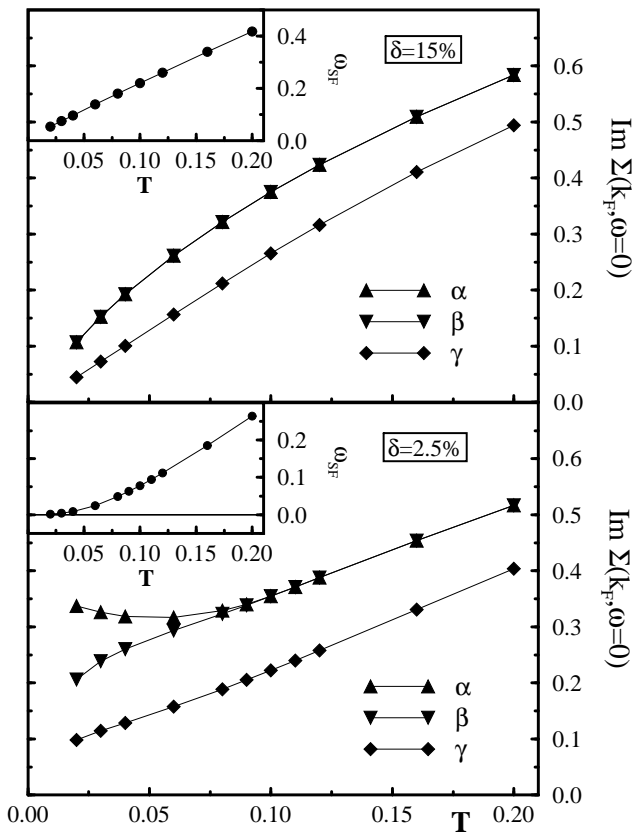}}}
\end{picture}
\caption{Im $\Sigma({\bf k}_F,\omega\! =\! 0)$ for three points on the
Fermi line (at maximal scattering rate, i.e. ${\bf k}_{F\alpha}$, as well
as ${\bf k}_{ F\beta}$, and ${\bf k}_{F\gamma}$ as defined in the caption of
Fig. 2) and hole dopings of $15 \%$ and $2.5\%$ in the temperature range
$T=[0.02t;0.2t]$. The insets display the temperature dependence of the spin
fluctuation energy scale $\omega_{sf}$.}
\label{fig3}
\end{figure}

Fig. \ref{fig3} summarizes the temperature dependence of the imaginary part 
of the on--shell ($\omega=0$) self--energy at the FS for two doping 
concentrations $\delta=15$\% and $2.5$\% for Fermi momenta on the BZ boundary,
the BZ diagonal, and the momentum of the maximal scattering rate, i.e. 
${\bf k}_{F\beta}$, ${\bf k}_{F\gamma}$, and ${\bf k}_{F\alpha}$, respectively.
While for $\delta=15$\% the momenta ${\bf k}_{F\alpha}$ and ${\bf k}_{F\beta}$ 
coincide at all temperatures depicted they {\em separate} below a critical 
temperature $T^\star$ for $\delta=2.5$\%. Therefore, the 
'$\alpha$--self--energy' in Fig.\ref{fig3}b) contains both, an explicit 
temperature dependence as well as an implicit one which is linked to the 
movement with temperature of ${\bf k}_{F\alpha}(T)$ along the FS line.

Two conclusions can be drawn from Fig. \ref{fig3}. First, we observe an
intimate relation between the temperature $T^\star(\delta)$ and the spin
fluctuation frequency $\omega_{sf}(T)$ which we define by the frequency of
the maximum in Im $T_{sf}({\bf q},\omega)$ at ${\bf q}={\bf Q}$. The insets
show the temperature dependence of $\omega_{sf}$ for both doping
concentrations. Hot spots start to form once the temperature drops {\em
below} the spin--fluctuation frequency scale, i.e. when
$T^\star(\delta)\sim\omega_{sf}(T)$. We find that this scenario is valid at
all doping levels which we have investigated and moreover that
$T^\star(\delta)$ increases with decreasing doping concentrations. It is
clearly tempting to relate $T^\star(\delta)$ with the pseudogap formation
temperature observed in ARPES \cite{Marshall,Loeser,Ding1,Ding2} and
optical conductivity experiments \cite{Basov}.

As a second consequence Fig. \ref{fig3} suggests the existence of a very small
energy scale which is manifest in the self--energy at low doping. At
$\delta=15$\% Im$\Sigma({\bf k}_{f},\omega=0)$ clearly extrapolates to zero for
vanishing temperatures at all FS momenta.  This is consistent with 
Fermi--liquid theory (FLQ). However, at $\delta=2.5$\% a similar behavior can 
not be anticipated. This pertains to all momenta irrespective of the additional
upturn of the scattering rate at ${\bf k}_{F\alpha}$ below $T^\star(\delta)$ 
which is partially due to the temperature dependent shift of 
${\bf k}_{F\alpha}(T)$. In order to recover low--temperature FLQ behavior we 
are forced to assume that Im $\Sigma({\bf k}_{F\alpha}(T),\omega=0,T)$ will 
approach zero below a second characteristic temperature $T^{FLQ}(\delta)$ below
$T^{\star}(\delta)$. Since $\omega_{sf}(T)$ is the only characteristic 
low--energy scale available it is natural to assume that $T^{FLQ}(\delta)\sim 
\omega_{sf}(T^{FLQ}(\delta))$. As is obvious from the inset of Fig. 
\ref{fig3}b) this temperature is expected to be very small for $\delta=2.5$\% 
and remains inaccessible within the numerical accuracy of our present 
computational scheme.

\begin{figure}
\begin{picture}(8.7,6.5)(1.5,10)
\put(0,0){\scalebox{0.59}{\includegraphics{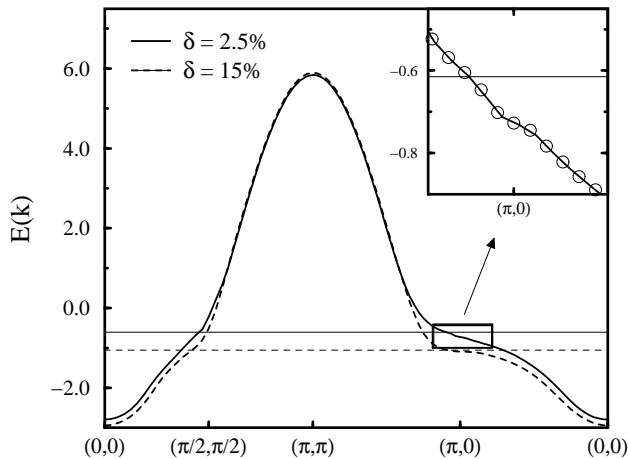}}}
\end{picture}
\caption{Quasi--particle dispersion relation for two different hole doping
  concentrations along the closed path in the BZ $(0,0)$ $\rightarrow$ 
  $(\pi,\pi)$ $\rightarrow$ $(\pi,0)$ $\rightarrow$ $(0,0)$. The horizontal 
  lines indicate the chemical potentials for both doping levels. The inset 
  shows a magnification of the vicinity of the $(\pi,0)$ point.}
\label{fig4}
\end{figure}

Finally we show the quasi--particle dispersion for $\delta=15$\% and $2.5$\% in
Fig. \ref{fig4}. The quasi--particle energies have been determined from the 
zeroes of the real part of the inverse Green function, i.e. from
$E({\bf k})=\epsilon({\bf k})-{\rm Re}\, \Sigma({\bf k},E({\bf k}))$. Far off 
the Fermi energy this figure seems to suggest a nearly rigid--band picture. 
However, close to the FS the situation is less trivial, in particular close to 
the wave vector ${\bf k}=(\pi,0)$, where quasi--particle energies get strongly
renormalized. With decreasing doping concentrations we find the flat band 
region around this wave vector to be 'pinned' to the Fermi energy and -- 
surprisingly -- deformed towards a stronger dispersion. While for 
$\delta=15$\% the quasi--particle energy $E(\pi,0)\simeq -1.09t$ is very close 
to the bare energy of $\epsilon(\pi,0)=-1.20 t$, the pinning gives rise to a 
substantial renormalization at $\delta=2.5\%$ where we find $E(\pi,0)\simeq 
-0.73t$. Qualitatively similar pinning, however quantitatively 
less pronounced, can be observed along the BZ diagonal. 

Focusing on the low--energy sector in the vicinity of the Fermi surface the 
inset in Fig. \ref{fig4} details yet another remarkable, though weak effect. 
While a continuous band dispersion is evident from this inset, with a well 
defined Fermi surface crossing on the BZ boundary, the quasi--particle energy 
develops a step like feature around $(\pi,0)$ for $\delta=2.5\%$. This is a 
precursor of the energy gap opening at half--filling and induces a pseudogap 
structure in the density of states. We have verified this feature to be robust 
against finite size scaling on up to $128\times 128$ lattices. Clearly the
magnitude of the pseudogap is too small to account for the recent
corresponding findings of ARPES \cite{Marshall,Loeser,Ding1,Ding2}. A
tempting remedy for this deficiency is to decrease the ratio of
$|t'/t|$ and thereby enhance the effects of nesting. Indeed, within
our FLEX scheme, we found that reducing $t'$ leads to a pronounced
increase of the pseudogap structure and moreover to the appearance of
a split--off precursor band similar to that observed in ARPES
\cite{JAWtoApp}. This is also consistent with QMC data obtained on the
pure $t$--model \cite{Hanke}. However, we believe that this route to
pseudogap behavior is inconsistent, both with the Fermi surface
topology of the cuprate materials as well as with the strong
anisotropy of the quasi--particle scattering rate, which both require
the inclusion of a finite hopping amplitude $t'$. Therefore a proper
understanding of the combined effects of anisotropy and pseudogap
formation requires further investigations.

In conclusion we have studied the anisotropic electronic properties of the
two--dimensional $U-t-t'$ Hubbard model which result from the unique
combination of the shape of the Fermi surface and strong scattering from
spin fluctuations. With decreasing hole doping and lowering of the
temperature hot spots develop on the  Fermi surface with
significantly enhanced scattering rates offering a physical explanation for
the observed strong anisotropies of the ARPES signals in the underdoped
cuprates. For realistic values of the ratio $t'/t$ clear Fermi surface
crossings are still maintained in our FLEX calculation and therefore a
pronounced pseudogap formation is not obtained. An extension to bilayer
type Hubbard models is a possible path for further study since the bilayer
splitting into a bonding and an antibonding band provides another important
low energy scale which may be of particular relevance near the $(\pi,0)$
point of the BZ.

This research was performed within the program of the Sonderforschungsbereich 
341 supported by the Deutsche Forschungsgemeinschaft (DFG).

\end{document}